# Optically enhanced coherent transport in $YBa_2Cu_3O_{6.5}$ by ultrafast redistribution of interlayer coupling

W. Hu[1*], S. Kaiser[1*], D. Nicoletti[1*], C.R. Hunt[1,4*], I. Gierz[1], M. C. Hoffmann[1], M. Le Tacon[2], T. Loew[2], B. Keimer[2], and A. Cavalleri[1,3]

[1] *Max Planck Institute for the Structure and Dynamics of Matter, Hamburg, Germany*
[2] *Max Planck Institute for Solid State Research, Stuttgart, Germany*
[3] *Department of Physics, Oxford University, Clarendon Laboratory, Oxford, United Kingdom*
[4] *Department of Physics, University of Illinois at Urbana-Champaign, Urbana, Illinois, USA*
* *These authors contributed equally to this work.*

**Nonlinear optical excitation of infrared active lattice vibrations[1] has been shown to melt magnetic[2] or orbital[3] orders and to transform insulators into metals[4,5]. In cuprates, this technique has been used to remove charge stripes and promote superconductivity[6], acting in a way opposite to static magnetic fields[7,8,9]. Here, we show that excitation of large-amplitude apical oxygen distortions in the cuprate superconductor $YBa_2Cu_3O_{6.5}$ promotes highly unconventional electronic properties. Below the superconducting transition temperature ($T_c$ = 50 K), inter-bilayer coherence is transiently enhanced at the expense of intra-bilayer coupling. Strikingly, even above $T_c$ a qualitatively similar effect is observed up to room temperature, with transient inter-bilayer coherence emerging from the incoherent ground state and similar transfer of spectral weight from high to low frequency. These observations are compatible with previous reports of an inhomogeneous[10] normal state that retains important properties of a superconductor[11,12,13,14], in which light may be melting competing orders[9,15] or dynamically synchronizing the interlayer phase. The transient redistribution of coherence discussed here could lead to new strategies to enhance superconductivity in steady state.**

Superconductors at equilibrium display two characteristic physical properties: zero DC resistance and the expulsion of static magnetic fields. The first of these properties manifests itself as a zero-frequency delta function in the real part of the optical conductivity $\sigma_1(\omega)$ and by a positive imaginary part $\sigma_2(\omega)$ that diverges at low frequency as $1/\omega$.

In high-$T_c$ cuprates, the layered structure gives rise to additional c-axis excitations of the superfluid, with the notable appearance of one or more longitudinal Josephson plasma modes due to tunneling of Cooper pairs between capacitively coupled superconducting planes.

In bi-layer cuprates, two longitudinal Josephson plasma modes are found[16,17], reflected by two peaks in the energy loss function -Im$[1/(\varepsilon_1(\omega)+i\varepsilon_2(\omega))]$. Within each family of cuprates, the longitudinal mode frequency quantifies the strength of the Josephson coupling between pairs of $CuO_2$ layers (Fig.1a). In addition, a peak in the real part of the conductivity, the so-called transverse Josephson plasma mode[18] (Fig. 1a), is observed at $\omega = \omega_T$[19,20,21,22,23,24,25]. This second dissipationless excitation of the superfluid[26] is characterized by simultaneous out of phase oscillations of the Josephson plasma within and between pairs of layers, and shares spectral weight with the zero-frequency conductivity peak.

In the specific case of $YBa_2Cu_3O_{6.5}$, the two longitudinal Josephson plasma modes appear as reflectivity edges and as peaks in the loss function near 30 and 475 cm$^{-1}$ (Fig.1 b2 and b4). The transverse plasma mode (Fig.1 b3) is observed near 400 cm$^{-1}$, and it is strongly coupled[27] to a 320-cm$^{-1}$ phonon from which it gains oscillator strength with decreasing temperature.

Here, we measure the transient c-axis THz-frequency optical properties of $YBa_2Cu_3O_{6.5}$ after excitation with mid-infrared optical pulses, both below and above $T_c$. Mid-infrared pump pulses of ~300 fs duration, polarized along the $c$ direction and tuned to 670 cm$^{-1}$ frequency (~15-µm, 83 meV, ± 15%), were made resonant with the infrared-active distortion shown in figure 1c. The 15-µm-wavelength pulses were generated by difference-frequency mixing in an optical parametric amplifier and focused onto the samples with a maximum fluence of 4 mJ/cm², corresponding to peak electric fields up to ~3 MV/cm. At these strong fields, the apical oxygen positions are driven in an oscillatory fashion by several percent of the equilibrium unit-cell distance (see supplementary material S5).

For temperatures below and immediately above $T_c$, where the largest changes of the optical properties were observed, we interrogated the solid with broadband THz probe pulses generated by gas ionization, covering the 20 and 500 cm$^{-1}$ frequency range. For temperatures far above $T_c$, smaller conductivity changes could be measured with sufficient signal to noise ratio only by using narrower-band pulses (20 – 85 cm$^{-1}$), generated by optical rectification in ZnTe.

The equilibrium low-frequency imaginary part of the optical conductivity $\sigma_2(\omega)$ (Fig. 1b1) is positive and increasing at frequencies below 30 cm$^{-1}$. Note that because of the contribution from normal transport by non-condensed quasi-particles, the overall equilibrium $\sigma_2(\omega)$ does not exhibit the 1/ω frequency dependence of a London superconductor. Such 1/ω frequency dependence can only be observed in this frequency range by measuring differential conductivity $\Delta\sigma_2(\omega) = \sigma_2(\omega, T < T_c) - \sigma_2(\omega, T > T_c)$.

The broadband photo-induced response of the superconducting state of YBa$_2$Cu$_3$O$_{6.5}$ ($T$ = 10 K) is reported in figures 2 and 3. The frequency- and time-delay-dependent optical properties were extracted from measurements of the amplitude and phase of the reflected electric field after photo-excitation, using the equilibrium optical properties of the material[21] (see Fig.1b) and taking into account the pump-probe penetration depth mismatch (see supplementary information S2). Immediately after excitation, a strong increase in the slope of σ$_2$(ω) was observed. Since the superfluid density at equilibrium is quantified as $\omega\sigma_2(\omega)|_{\omega\to 0}$, the increase in the slope of σ$_2$(ω) suggests a transient enhancement of the superfluid density of the superconductor (see upper panel of Fig. 2a). The frequency-dependent imaginary conductivity is shown for all pump-probe time delays in the color plot.

In Fig. 2b, we report the corresponding changes in the inter and intra-bilayer coupling by plotting time- and frequency-dependent energy-loss function -Im[1/(ε$_1$(ω,τ)+iε$_2$(ω,τ))]. The 30 cm$^{-1}$ peak, which reflects the inter-bilayer longitudinal plasma mode at equilibrium, reduces in amplitude after photo excitation, as a second higher-frequency peak appears at 60 cm$^{-1}$. Simultaneously, the intra-bilayer peak, which at equilibrium is observed at 475 cm$^{-1}$, shifts to the red. All transient shifts in the loss function relax back to the equilibrium spectrum with a 7 ps exponential time constant (black dashed line in Fig. 2b).

Finally, figure 3 displays the corresponding dynamics of the real part of the conductivity σ$_1$(ω,τ). At equilibrium, σ$_1$(ω) nears zero below 80 cm$^{-1}$, with several phonon peaks between 100-300 cm$^{-1}$. These phonon peaks remain virtually unaffected and only a small increase in σ$_1$(ω) is detected below 80 cm$^{-1}$. The strongest light-induced changes in σ$_1$(ω) are found at high frequency, where the transverse plasma mode at 400 cm$^{-1}$ shifts to the red. Note that the red shift of this mode (σ$_1$(ω) peak at 400 cm$^{-1}$) is consistent with the red shift of the loss function peak at 475 cm$^{-1}$. The transverse mode frequency follows $\omega_T^2 = (d_2\omega_{Jp1}^2 + d_1\omega_{Jp2}^2)/(d_1 + d_2)$, where $d_1$ and $d_2$ are the thickness of the inter- and intra-bilayer junctions, respectively, and $\omega_{Jp1}$ and $\omega_{Jp2}$ are the corresponding Josephson plasma frequencies. Because $\omega_{Jp2} \gg \omega_{Jp1}$, the change in transverse plasma mode position $\omega_T$ is dominated by $\omega_{Jp2}$.

To analyze the photo-induced dynamics below $T_c$, we first note that the changes in optical properties are only partial. For example, the light induced 60-cm$^{-1}$ loss function peak (figure 2b) takes only a fraction of the equilibrium spectral weight at 30 cm$^{-1}$. This is interpreted a signature of an inhomogeneous light induced phase, in which only a fraction of the equilibrium superconducting state is being transformed, a physical situation that can be well described by Bruggeman's effective medium model[28]

$$f \frac{\tilde{\varepsilon}_T(\omega) - \tilde{\varepsilon}_E(\omega)}{\tilde{\varepsilon}_T(\omega) + 2\tilde{\varepsilon}_E(\omega)} + (1-f) \frac{\tilde{\varepsilon}_S(\omega) - \tilde{\varepsilon}_E(\omega)}{\tilde{\varepsilon}_S(\omega) + 2\tilde{\varepsilon}_E(\omega)} = 0$$

The effective dielectric function $\tilde{\varepsilon}_E(\omega)$ is determined here by the dielectric function $\tilde{\varepsilon}_T(\omega)$ of the photo-transformed regions, which occupy a volume fraction $f$, and by the dielectric function $\tilde{\varepsilon}_S(\omega)$ of the remaining (1-f) volume, which we assumed to retain the properties of the equilibrium superconducting state.

The transient optical properties at all frequencies could then be fitted using a minimum number of free parameters. We considered the dielectric function $\tilde{\varepsilon}_T(\omega)$ of a second superconductor with similar optical properties to YBa$_2$Cu$_3$O$_{6.5}$ at equilibrium, but with different values of $\omega_{Jp1}$ and $\omega_{Jp2}$ and with a different 320 cm$^{-1}$ phonon width (see supplementary material S3 for details). In addition, the filling fraction $f$ was left as free parameter. All transient features were well reproduced by this fit, with a maximum transformed volume fraction of $f$ = 20%.

From the effective medium fit, we extract the unscreened Josephson plasma frequency for the perturbed volume fraction *f*. The unscreened inter-bilayer Josephson plasma frequency increases from $\omega_{Jp1}$ = 110 to 310 cm$^{-1}$, and the unscreened intra-bilayer Josephson plasma frequency decreases from $\omega_{Jp2}$ = 1030 to 950 cm$^{-1}$, conserving the total coherent weight, which scales with $\omega_{JP1}^2 + \omega_{JP2}^2$ (Note that the position of the Josephson plasma edges in the reflectivity, and the peaks in the loss function, are located at frequencies much smaller than $\omega_{Jp1}$ and $\omega_{Jp2}$. Here the edge/peak positions are determined by the screened plasma frequency $\widetilde{\omega_{Jp}} = \omega_{Jp}/\sqrt{\varepsilon_\infty}$, with $\varepsilon_\infty$= 4.5. The interband contribution shifts the peak/edge positions further to even lower frequencies than $\widetilde{\omega_{Jp}}$).

We next turn to the response immediately above $T_c$, for which the signal was still large enough to allow for measurements with the same broadband source. Fig. 4a shows the equilibrium and light-induced σ$_2$(ω) for *T*=60 K (> $T_c$ = 50 K). We observe a short-lived enhancement of the low frequency σ$_2$, which becomes positive and increases with decreasing frequency. Note that this transient σ$_2$(ω) is very similar to that of the equilibrium *superconductor* at 10 K (grey line in Fig.2a). Figure 4b displays the corresponding loss function. The appearance of a broad loss function peak at 55 cm$^{-1}$, absent in the normal state, reflects the emergence of a longitudinal plasma mode[29] approximately at the same frequency where the blue shifted mode was observed below $T_c$. Furthermore, the 475 cm$^{-1}$ loss function peak (Fig.4b) and the 400 cm$^{-1}$ peak in σ$_1$ (Fig.4c) both shift to the red, identifying a clear analogy between the below $T_c$ and above $T_c$ data. These broadband data at 60 K were also successfully fitted with the same effective medium model, where the unscreened inter-bilayer plasma frequency increases from $\omega_{Jp1}$= 0 cm$^{-1}$ to $\omega'_{Jp1}$=250 cm$^{-1}$, and the unscreened intra-bilayer

Josephson plasma frequency $\omega_{Jp2}$ =1030 cm$^{-1}$ reduces to $\omega'_{Jp2}$ =960 cm$^{-1}$ (see Supplementary Materials S3), again conserving the total coherent weight $\omega_{JP1}^2 + \omega_{JP2}^2$. The response over a broader range of temperatures below and above $T_c$ was measured with narrowband THz pulses, which were more sensitive to smaller changes in optical properties. Figure 5 displays three representative sets of results at 10 K, 100 K and 300 K, as extracted from the experiment using the same procedure as for the broadband data. The 10 K results confirm the photo-induced enhancement in $\sigma_2(\omega)$ (Fig.5a1) and a weak increase in $\sigma_1(\omega)$ (Fig.5a2). For comparison, the differential measurement of the red-shift in $\sigma_1(\omega)$ near 400 cm$^{-1}$ (extracted from the broadband measurements of Fig. 3) is displayed in Fig.5a3.

Above $T_c$, at 100 K (Fig.5b1-5b3) we also observed an increase and change in $\sigma_2(\omega)$, which becomes positive below 30 cm$^{-1}$, as already shown for the 60 K data of Fig.4a. The corresponding changes in the low frequency $\sigma_1(\omega)$ are negligible (Fig.5b2), and a transfer of spectral weight from the 400 cm$^{-1}$ mode to lower frequencies is detected. At 300 K (Fig. 5c1-5c3), smaller changes with similar qualitative characteristics are observed. The conductivity response discussed in Fig.5a-c is complemented by plots of reflectivity and loss function, reported in Fig.5d-f. These figures evidence the appearance of a clear longitudinal plasma mode at ~ 60 cm$^{-1}$. By fitting the temperature-dependent photo-induced enhancement $\omega\Delta\sigma_2|_{\omega\rightarrow 0}$ with an empirical mean-field law of the type $\propto \sqrt{1 - \frac{T}{T'}}$, a temperature scale $T' = 310 \pm 10$ K is extracted for the disappearance of the effect (Fig.5g).

We next turn to a critical discussion of all the experimental results reported above. Below $T_c$, we observe strengthening of the low-frequency inter-bilayer coupling, occurring at the expense of that within the bilayers. A change in the Josephson coupling strength in cuprates may be ascribed to more than one physical origin. For example, the dynamical coupling between the layers may change because the charging energy of the planes is modified. However, if this were the case, both plasma modes below $T_c$ should shift in the same direction, that is to lower/higher frequencies if the electronic compressibility[18] increased/decreased. Similarly, a reduction or increase of the interlayer coupling, either by ionization of the condensate across the gap or by an increase in the total number of Cooper pairs, would lead to a shift of the longitudinal modes in the same direction, either to the red or to the blue. As $\omega_{JP1}^2 + \omega_{JP2}^2$ is constant throughout the dynamics, we conclude that the light only rearranges the relative tunneling strengths, with coherence being transferred from the bilayers to the inter-bilayer region. This is further supported by a partial sum rule analysis for $\sigma_1(\omega)$ over the measured spectral range. We compare the reduction in the spectral weight at finite frequencies (20-500 cm$^{-1}$), which is dominated by the $\omega_T$ peak, with the enhancement at low frequency. The zero frequency peak, proportional to the superfluid density, cannot be measured directly in $\sigma_1(\omega)$ but can be quantified either by $\omega\sigma_2(\omega)$ for $\omega \to 0$ or, equivalently, as $\omega_{JP1}^2 \omega_{JP2}^2 / \omega_T^2$ [23,27]. Since $\omega_{JP2}^2 \approx \omega_T^2$, the c-axis superfluid density is approximately proportional to $\omega_{JP1}^2$, which increases after photo-excitation, thus indicating an enhancement of the superfluid density. The spectral weight loss in the light-induced state, computed as $\frac{120}{\pi} \int_0^{\omega_m} (\sigma_{1T} - \sigma_1^{Equilibrium}) d\omega = -1.0 \times 10^5 \ cm^{-2}$ (here $\sigma_{1T}$ is the optical conductivity of the photo-perturbed region, and $\omega_m = 500$ cm$^{-1}$ is the cut-off frequency, which is the highest frequency we could access experimentally) is comparable to the enhancement of the inter-bilayer coherence $\Delta\omega_{JP1}^2 = 8.4 \times 10^4 \ cm^{-2}$ of the same photo-perturbed region. Thus, the weakening of the intra-bilayer coupling alone seems to be responsible for the observed enhancement in superfluid density. Similar to the below $T_c$ case, the finite-energy sum rule for the broadband data above $T_c$ (at $T$ = 60 K) shows the spectral weight loss in $\sigma_1$ is $7 \times 10^4 \text{cm}^{-2}$, and the emergence of inter-bilayer coherence is $\Delta\omega_{Jp1}^2 = 6 \times 10^4 \text{cm}^{-2}$ for the photo-perturbed region ($f$ = 19%).

Our effective medium analysis also shows that the phonon at 320 cm$^{-1}$, which is strongly coupled to the transverse Josephson plasma mode, is sharpened after excitation (see Supplementary Materials S3). Thus, rearrangement of the lattice may explain the transfer of coupling strengths between the two "junctions". In the same spirit of the pair-density wave interpretation of quenched Josephson coupling in stripe-ordered cuprates[30], which posits disruptively interfering tunneling between coupled planes, one explanation of the present data may be that the relative strength of inter and intra-bilayer coupling at equilibrium is affected by interference effects caused by charge order in the planes. If charge order were perturbed at constant superfluid density, one tunneling strength may increase at the expense of the other.

In the above $T_c$ broadband data, a qualitative similarity to the spectral redistributions observed below $T_c$ is found. Most conservatively, the experimental data could be fitted by the optical properties of a Drude metal with a very long scattering time $\tau_s \sim 7$ ps (corresponding to the lifetime of the state, see also Supplementary Materials S2). This value of $\tau_s$ implies a DC mobility $\mu = \frac{e\tau}{m} \sim 10^3 - 10^4 \frac{\text{cm}^2}{\text{V}\cdot\text{s}}$ (depending on the carrier effective mass). Note that such high mobility would be highly unusual for incoherent transport in oxides. Furthermore, as the position of the edge does not move with the number of absorbed photons and only the fraction of material that is switched is a function of laser field, the results are hardly compatible with above-gap photoconductivity.

A more exotic effect that may give rise to extraordinarily high mobility and an anomalous dependence on the laser field may be conduction by a sliding one-dimensional charge density wave (CDW)[31]. If a non-commensurate CDW in this density range[32] were to become de-pinned when the lattice is modulated, and if such a wave could slide along the c axis, it may pin again only a few picoseconds after excitation. Yet, it appears unlikely that the same carrier density as in the equilibrium superconducting state should contribute to such a CDW conductor.

An interpretation based on transient superconducting coherence induced far above $T_c$, is in our view the most plausible. First, the photo-induced change in the imaginary conductivity $\Delta\sigma_2(\omega)$ tracks very well the change $\Delta\sigma_2(\omega) = \sigma_2(\omega, 10\ \text{K}) - \sigma_2(\omega, 60\ \text{K})$ measured at equilibrium when cooling below $T_c$ (see inset in figure 5b1). Secondly, the photo-induced plasma edge is very close to the equilibrium inter-bilayer Josephson Plasma Resonance, showing that light induced transport above $T_c$ involves a density of charge carriers very similar to the density of Cooper pairs that tunnel between the planes in the equilibrium superconductor.

Transient superconducting coherence could not be caused by quasi-particle photo-excitation[33,34], which was shown in the past to increase $T_c$ at microwave[35,36,37,38,39] or optical[40,41] frequencies, as the enhancement was only observed here when the pump was tuned to the phonon resonance (see Supplementary Material S4). Rather, the nonlinear excitation of the lattice may create a displaced crystal structure[42] with atomic positions more favorable to high temperature superconductivity, for example "melting" an ordered state that competes with superconductivity or cause a displacement of the apical oxygen away from the planes[43,44,45,46,47]. Similarly to what discussed for the below $T_c$ data, the entire gain of coherent spectral weight above $T_c$ can be accounted for by considering the red shift of the transverse plasma mode near 400 cm$^{-1}$. The appearance of this mode above equilibrium $T_c$ has been discussed in the past as a signature of residual intra-bilayer superfluid density[48] in the normal state, and we speculate that redistribution of intra-bilayer coherence may explain our data.

We also mention the possibility that the observed effects result from dynamical stabilization[49,50,51] of superconducting phases. As the 15-µm modulation used here occurs at frequencies high compared to plasma excitations between planes, one could envisage a dynamically-stabilized stack of Josephson junctions[52], by direct coupling of the oscillatory field to the order parameter.

In summary, we showed that light stimulation redistributes interlayer Josephson coupling in the superconducting state of bi-layer YBa$_2$Cu$_3$O$_{6.5}$, enhancing inter-bilayer coupling at the expense of the coupling within the bilayers. Above $T_c$, a similar phenomenology is observed, including a positive dynamical inductance, a reflectivity edge and a redistribution of spectral weight from high to low frequencies. The hypothesis of transient superconducting coupling surviving to room temperature, would imply that pre-existing coherence is redistributed. This latter scenario poses stringent constraints on our understanding of the normal state, and may lead to strategies for the creation of higher temperature superconductivity over longer timescales, in driven steady state or even by designing appropriate crystal structures.

**Acknowledgements**

The authors are grateful to J. Orenstein, S. Kivelson, D. Basov, D. van der Marel, C. Bernhard, A. Leitenstorfer, and L. Zhang for extensive discussions, for their many suggestions and advice on the data analysis. Technical support from J. Harms and H. Liu are acknowledged.

**Author contributions**

A. C. conceived the project. W. H., I. G. performed the measurements with broadband gas source. D. N., C. R. H., and S. K. performed the narrowband measurements. W. H., C. R. H., D. N. and I. G. analyzed the data and discussed results with all authors. W. H. built the mid-infrared pump-broadband THz probe setup with the support from M. C. H. The MIR pump-narrowband THz probe setup was built by S. K. and D. N. $YBa_2Cu_3O_{6.5}$ single crystals were synthesized by T. L., with guidance from M.L.T. and B. K. The manuscript was written by A. C. together with W. H. and S. K., and with input from all authors.

**Competing financial interests**
The authors declare no competing financial interests.

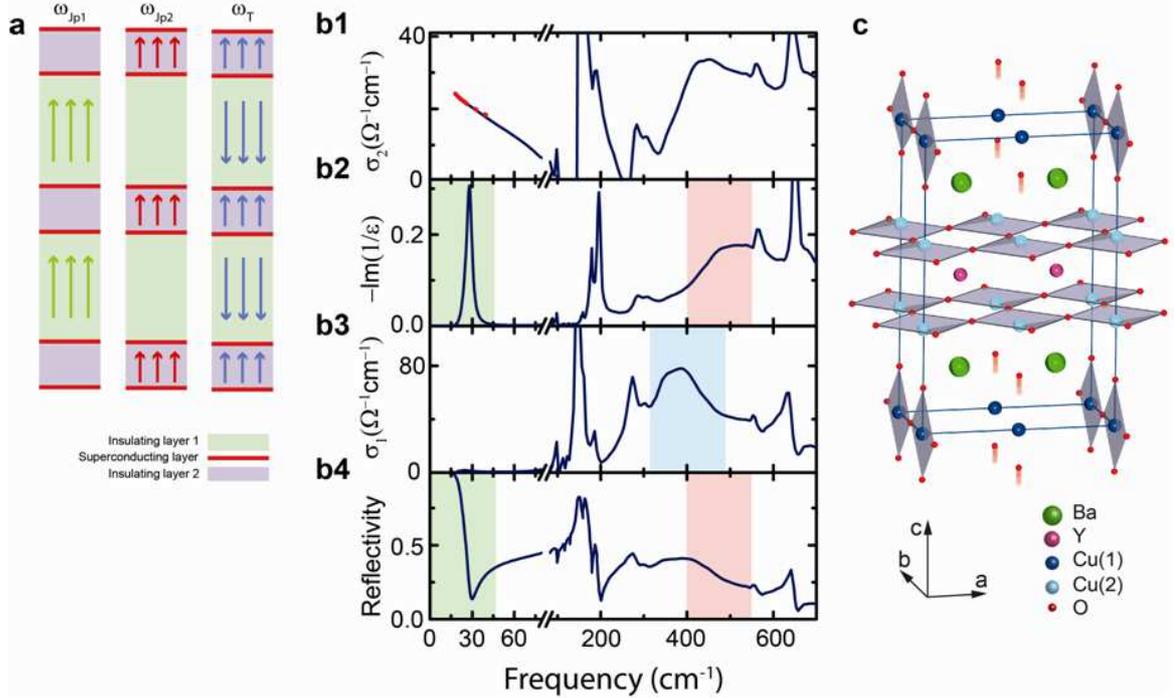

**Figure 1. Bilayer structure and c-axis optical features in superconducting YBa$_2$Cu$_3$O$_{6.5}$. (a).** In the superconducting state, the structure of YBa$_2$Cu$_3$O$_{6.5}$ can be viewed as two Josephson junctions in series, which gives rise to two longitudinal modes ($\omega_{Jp1}$, $\omega_{Jp2}$) and a transverse mode ($\omega_T$) (arrows indicate the direction of the current[18]). **(b1-b4).** Equilibrium c-axis optical properties for YBa$_2$Cu$_3$O$_{6.5}$[21] (the low frequency Josephson plasma edge was characterized by narrow-band THz spectroscopy. See Supplementary Material S1). Superconductivity is evidenced by the 1/ω divergence (red dashed-dotted line) in the imaginary part of the optical conductivity, σ$_2$(ω). Two longitudinal Josephson plasma modes appear as two peaks in the loss function, -Im(1/ε), and two edges in the reflectivity (~30 cm$^{-1}$, ~475 cm$^{-1}$, shaded area). The transverse plasma mode appears as a broad peak around 400 cm$^{-1}$ in the real part of the optical conductivity σ$_1$(ω). **(c)** Structure of YBa$_2$Cu$_3$O$_{6.5}$[53] and sketch of the optically driven distortion for the apical oxygen. Two conducting CuO$_2$ planes are separated by Y atoms (pink) and form a bilayer unit. Ba atoms (green) and the CuO$_4$ ribbons (Cu(1), and O in the *bc*-plane) separate bilayer units. The excitation of the infrared-active B$_{1u}$ mode at ~15 μm (670 cm$^{-1}$) displaces the apical oxygen atoms along the c direction[54].

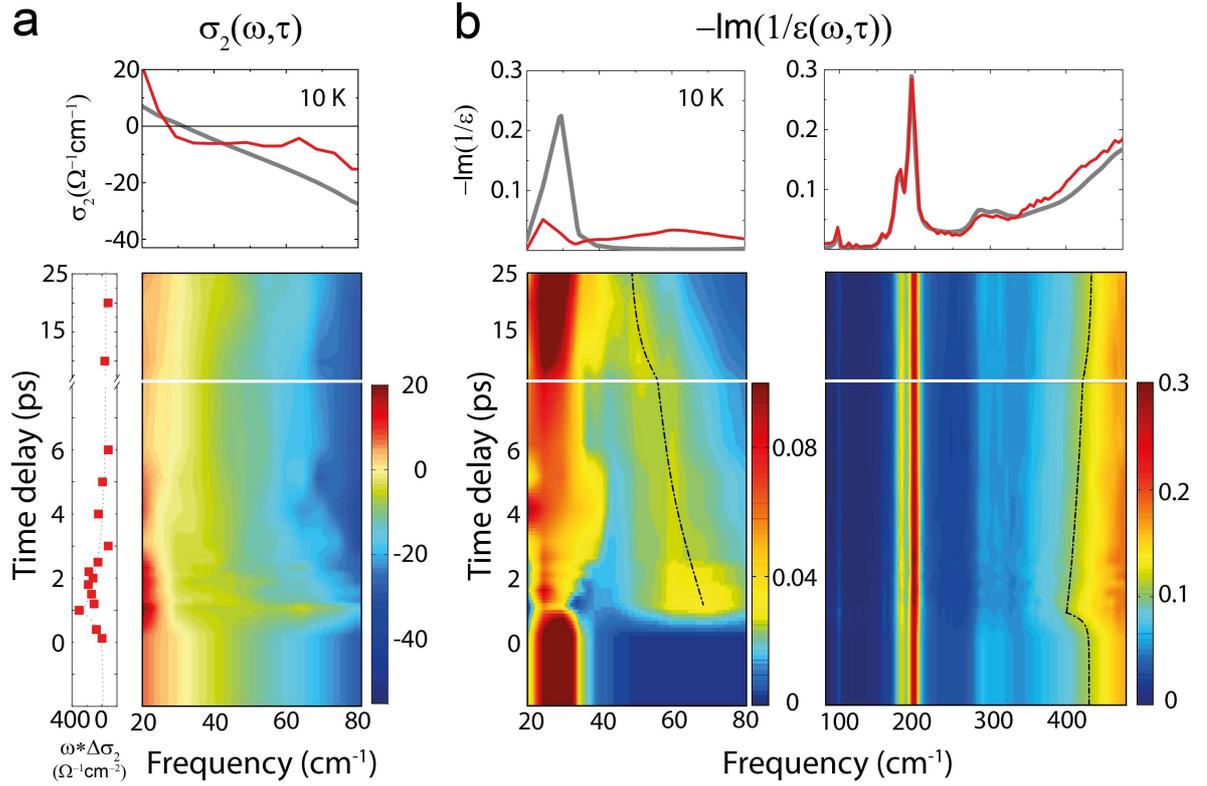

**Figure 2. Transient imaginary conductivity at low frequencies (< 80 cm$^{-1}$) and broadband (20-500 cm$^{-1}$) loss function response at 10 K ($T < T_c$). (a)** Imaginary part of the optical conductivity $\sigma_2(\omega,\tau)$ and **(b)** the energy loss function $-\text{Im}(1/\varepsilon(\omega,\tau))$ of of YBa$_2$Cu$_3$O$_{6.5}$. **Upper panels:** transient optical constants at maximum response (red), and the equilibrium values (gray). **Lower panels:** transient optical constants as a function of frequency and time delay. The prompt increase of the $\sigma_2(\omega)$ slope is followed by a decay back to equilibrium within few picoseconds, as shown in the one dimensional plot on the left hand side of the color plot, reporting $\omega\Delta\sigma_2(\omega,\tau)$ for $\omega = 20$ cm$^{-1}$. A splitting of the low frequency inter-bilayer Josephson plasma mode (~30 cm$^{-1}$) and a red-shift of the high energy intra-bilayer Josephson plasma mode (~475 cm$^{-1}$) can be seen in the energy loss function. The decay of both the low and the high frequency plasma modes back to equilibrium follow the same exponential decay of 7 ps (dashed-dotted lines).

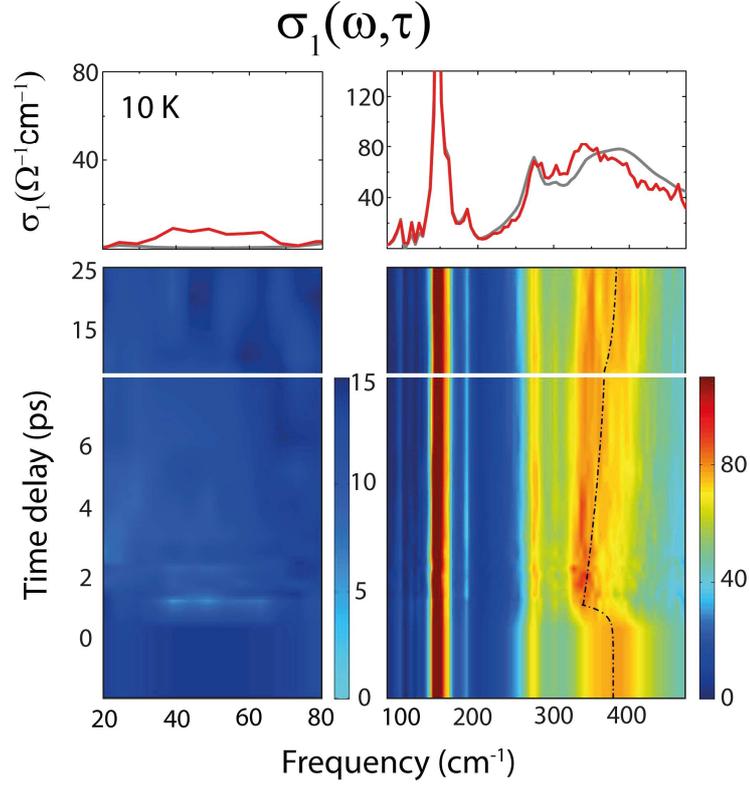

**Figure 3. Transient real part of the conductivity of YBa$_2$Cu$_3$O$_{6.5}$ at $T$ = 10 K ($T <$ $T_c$). Upper panels:** equilibrium (gray) and photo-excited real part of the optical conductivity $\sigma_1(\omega)$ at maximum transient response (red). A small contribution below 80 cm$^{-1}$ in the low frequency $\sigma_1(\omega)$, and a red-shift of the transverse plasmon around 400 cm$^{-1}$ are seen at maximum transient response. Lower panels: photo-excited $\sigma_1(\omega)$ at various time delays. The light-induced red-shift of the 400 cm$^{-1}$ mode follows the same 7 ps exponential decay (dashed-dotted line) as the transient changes in the two Josephson plasma modes in the loss function (figure 2b).

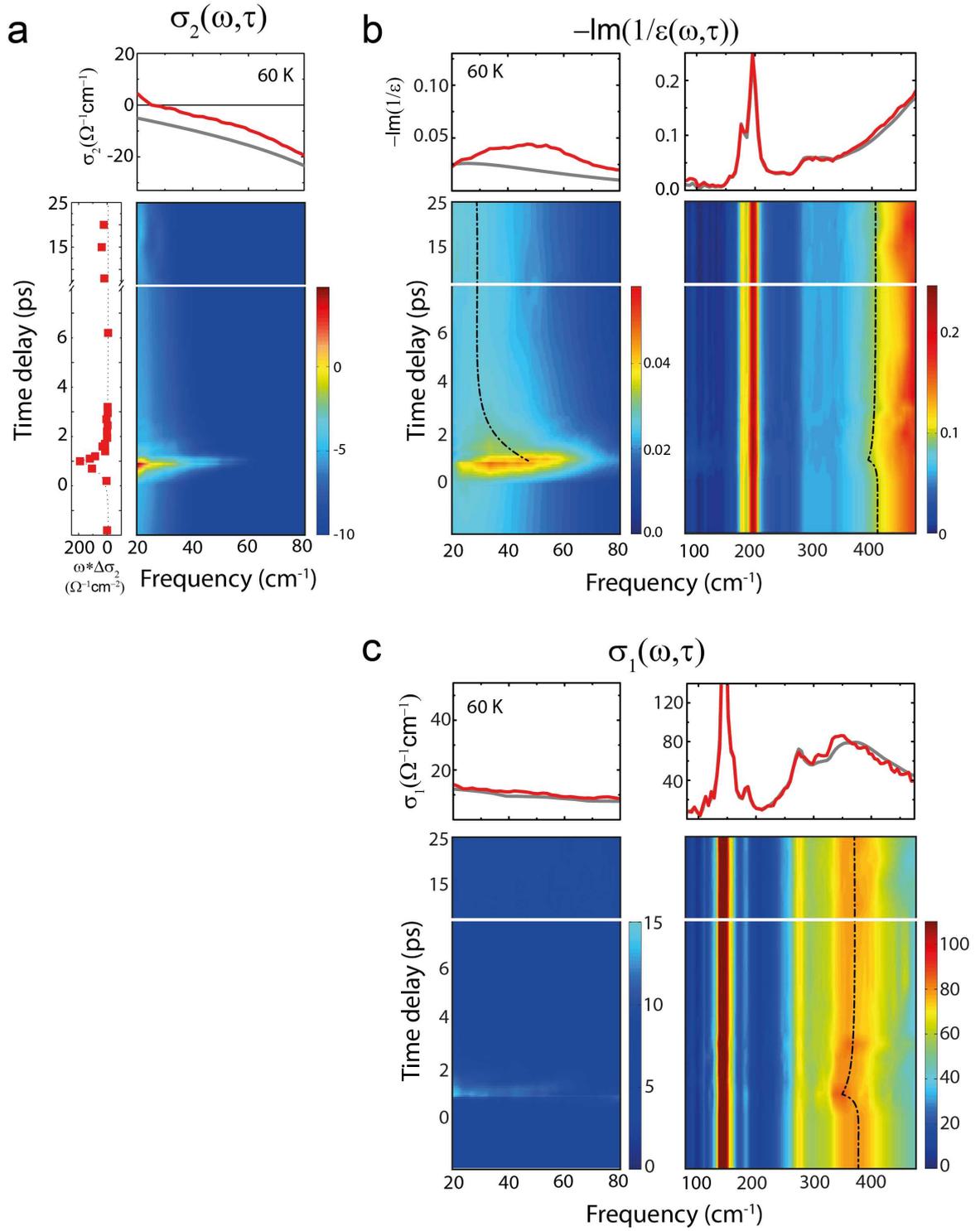

**Figure 4. Transient optical properties above $T_c$ (at $T$ = 60 K). (a)** Imaginary part of the optical conductivity $\sigma_2(\omega,\tau)$. An enhancement between 20 to 80 cm$^{-1}$, and a $1/\omega$-like divergence at low frequencies is seen in $\sigma_2$. **(b)** Loss function $-\text{Im}(1/\varepsilon(\omega,\tau))$. The equilibrium loss function is featureless below 80 cm$^{-1}$, and shows a strong peak at 475 cm$^{-1}$. Upon mid-infrared pumping, a peak at 55 cm$^{-1}$ develops, indicating photo-induced inter-bilayer coherence. At higher frequencies, a red-shift of the 475 cm$^{-1}$

intra-bilayer plasmon is seen. **(c)** Real part of the optical conductivity $\sigma_1(\omega,\tau)$. A red-shift of the 400 cm$^{-1}$ transverse plasma mode is observed. See Supplementary Materials S2 for details on the lifetime of the state above $T_c$.

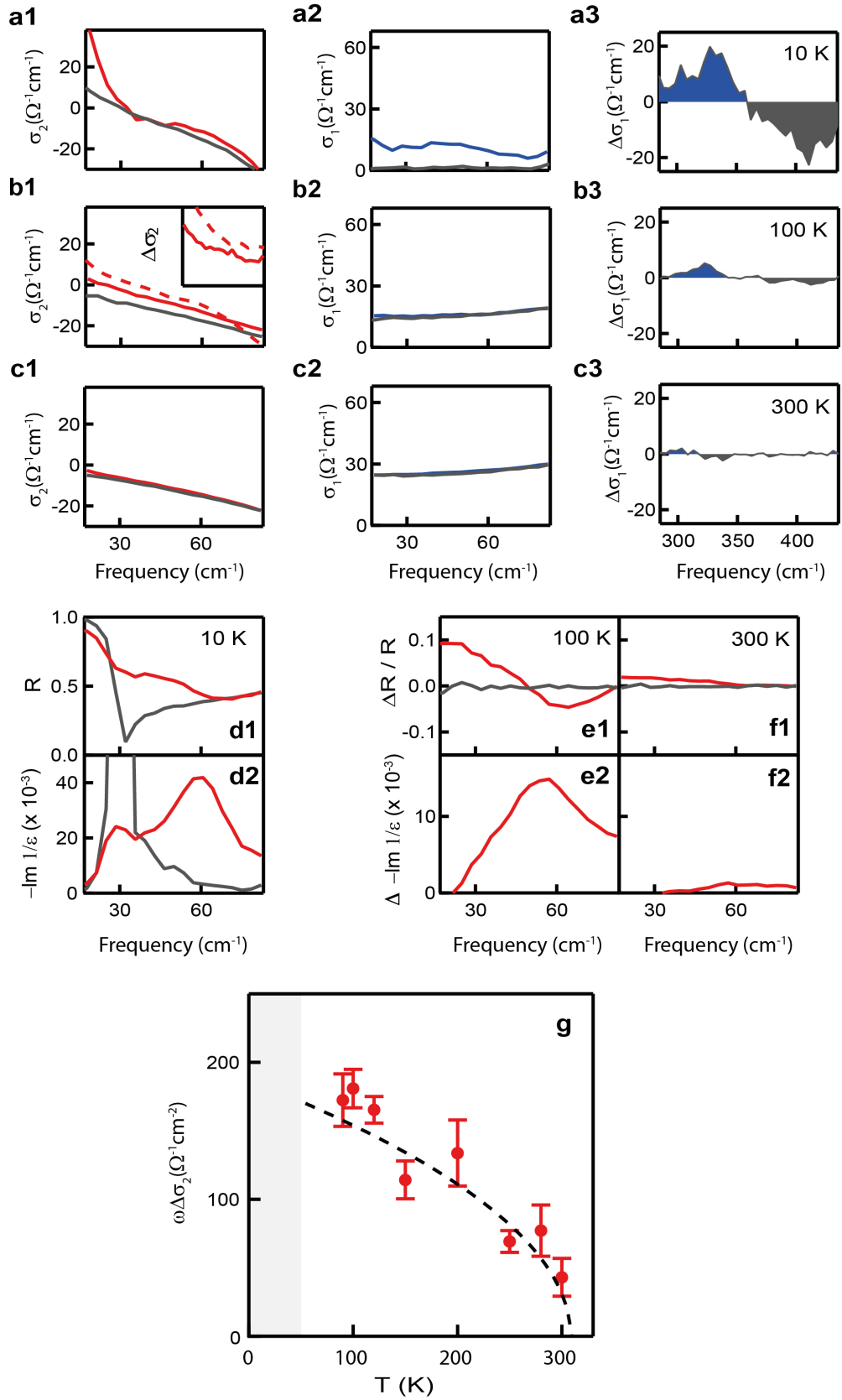

**Figure 5. Low frequency transient optical properties below and far above $T_c$**
**(a1-c1)** Red lines: Imaginary part of the conductivity $\sigma_2(\omega)$ at maximum transient response for 10 K ($T < T_c$), 100 K and 300 K. Grey lines: σ$_2$ at equilibrium. The red dashed line in **b1** shows the equilibrium σ$_2$ at 10 K (< $T_c$). The inset shows the transient change $\Delta\sigma_2 = \sigma_2(\omega, transient) - \sigma_2^0(\omega, equilibrium)$ (red line) compared with the equilibrium conductivity change by decreasing temperature: $\Delta\sigma_2(\omega, \Delta T) = \sigma_2^0(\omega, 10\,K) - \sigma_2^0(\omega, 60\,K)$ (red dashed line). **(a2-c2)** Blue lines: real part of the conductivity σ$_1$(ω) measured at maximum transient response for 10 K ($T < T_c$), 100 K and 300 K. Grey lines: equilibrium σ$_1$. **(a3-c3)** Photo-induced change Δσ$_1$ due to the red-shift of the transverse plasma mode. Blue/grey area: gain/loss of spectral weight. **(d1-d2)** Below $T_c$, the reduction of the equilibrium Josephson plasma mode can be found in the reflectivity (d1) and the loss function (d2), and a new plasma mode is seen at ~55 cm$^{-1}$. **(e1-f2)** Differential changes in reflectivity at 100 K (e1) and 300 K (f1). Changes in the loss function at 100 K (e2) and 300 K (f2). **(g)** Photo-induced enhancement $\omega\Delta\sigma_2|_{\omega\to 0}$ as a function of temperature. The dashed curve is a fit obtained with an empirical mean field dependence of the type $\propto \sqrt{1 - \frac{T}{T'}}$. The error bars indicate the standard deviation from the mean ω*Δσ$_2$(ω) in the region between 20 and 50 cm$^{-1}$.

**SUPPLEMENTARY INFORMATION**

# Optically enhanced coherent transport in YBa$_2$Cu$_3$O$_{6.5}$ by ultrafast redistribution of interlayer coupling

W. Hu[1*], S. Kaiser[1*], D. Nicoletti[1*], C.R. Hunt[1,4*], I. Gierz[1], M. C. Hoffmann[1], M. Le Tacon[2], T. Loew[2], B. Keimer[2], and A. Cavalleri[1,3]

[1] *Max Planck Institute for the Structure and Dynamics of Matter, Hamburg, Germany*
[2] *Max Planck Institute for Solid State Research, Stuttgart, Germany*
[3] *Department of Physics, Oxford University, Clarendon Laboratory, Oxford, United Kingdom*
[4] *Department of Physics, University of Illinois at Urbana-Champaign, Urbana, Illinois, USA*

**S1 Sample Growth and Characterization**

YBa$_2$Cu$_3$O$_{6+\delta}$ crystals of typical dimensions 2 x 2 x 1 mm$^3$ were grown in Y-stabilized zirconium crucibles[1]. The hole doping of the Cu-O planes was adjusted by controlling the oxygen content of the CuO chain layer δ by annealing in flowing O$_2$ and subsequent rapid quenching. A sharp superconducting transition at $T_c$ = 50 K (Δ$T_c$~2 K) was determined by *dc* magnetization measurements, as shown in Figure FS1.1A.

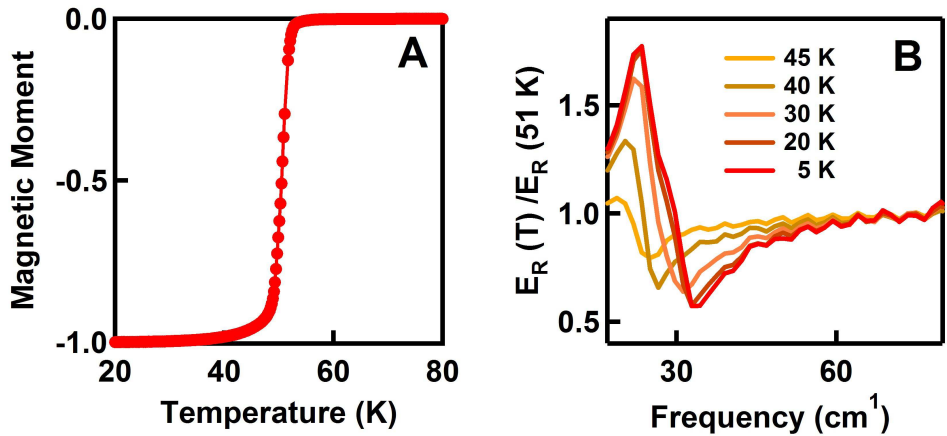

**Figure FS1.1 (A)** SQUID characterization of the dc magnetization across the superconducting transition. **(B)** Josephson plasma resonance in the relative THz reflected field ratios (below T$_c$ divided by above T$_c$) for YBa$_2$Cu$_3$O$_{6.5}$.

---

[1] S. I. Schlachter, U. Tutsch, W. H. Fietz, *et al.*, "Pressure effect and specific heat of RBa$_2$Cu$_3$O$_x$ at distinct charge carrier concentrations: Possible influence of stripes", Int. J. Mod. Phys. B **14**, 3673 (2000).

The equilibrium optical properties were characterized by THz time-domain spectroscopy. Single-cycle THz pulses (20-85 cm$^{-1}$) were focused onto the sample surface at 30° incidence, with polarization perpendicular to the Cu-O planes. The reflected electric field was measured by electro-optic sampling at different temperatures, below and above $T_c$.

Figure FS1.1B displays the absolute value of the frequency-dependent reflected electric field at a given temperature $|\tilde{E}_R(T)|$, normalized to the same quantity measured above $T_c$. Below $T_c$ a strong edge appears $\sim$ 30 cm$^{-1}$. This feature corresponds to the Josephson plasma resonance (JPR).

**S2 Derivation of the complex conductivity from differential reflectivity**

The transient optical properties were probed by broadband single cycle THz pulses generated in a laser-ionized plasma[2], covering a bandwidth from 20 to 500 cm$^{-1}$. The source was driven by 800 nm pulses with 1 mJ energy and 35 fs duration from a 2 kHz Ti:sapphire laser. The broadband THz pulses were polarized along the c-axis of the YBa$_2$Cu$_3$O$_{6.5}$ single crystal. The out-of-plane optical properties were measured at normal incidence. Mid-infrared pump pulses with a wavelength of 15 µm, 300 fs duration, and pulse energies of 8 µJ were generated by optical parametric amplification and subsequent difference frequency generation in GaSe[3].

The transient optical properties were also probed in reflection between 20 and 85 cm$^{-1}$ by narrowband THz pulses, which were generated by optical rectification of 800 nm femtosecond pulses from a 1 kHz Ti:sapphire laser in a ZnTe crystal. The reflected pulses were electro-optically sampled by 800 nm pulses in a second ZnTe crystal. For the frequency range where the two different THz probe setups overlap, a good agreement of the transient optical constants was found, with a better signal-to-noise ratio for the data obtained from the narrowband THz probe setup.

The pump-induced change in the reflected field was measured at each time delay $\tau$ during the dynamically evolving response of the material. For each $\tau$ the relative delay between excitation and sampling pulse was kept fixed, and the THz transient was scanned with respect to these two, changing the internal delay $t$. Therefore each point in the THz profile probed the material at the same time delay $\tau$ after excitation.

---

[2] I.-C. Ho, X. Guo, and X.-C. Zhang, "Design and performance of reflective terahertz air-biased-coherent-detection for time-domain spectroscopy," *Opt. Express* **18,** 2872 (2010).

[3] C. Manzoni, M. Först, H. Ehrke, et al, "Single-shot detection and direct control of carrier phase drift of midinfrared pulses," *Opt. Lett.* **35,** 757 (2010).

Note that with this method, the time resolution is not determined by the intensity envelope of the THz transient, which is generally frequency-chirped and can be > 1 ps long, but by the Fourier limit of the pulse, i.e. the bandwidth.

Figure FS2.1 shows a schematic of such a two-dimensional measurement, for which the time coordinate $\tau$ defines the instant in time of the sample dynamics and the time coordinate $t$ defines its spectral properties.

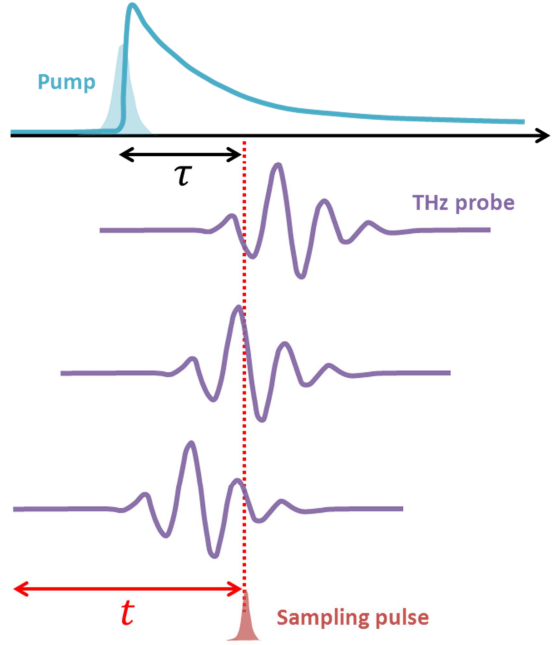

**Figure FS2.1** Measurement of the time and frequency dependent THz response at a single time delay $\tau$ during the materials dynamics. A mid-infrared pulse (blue) triggers the dynamics in the $YBa_2Cu_3O_{6.5}$ sample. The spectral response at each time delay $\tau$ is obtained by fixing the delay between pump pulse and electro-optic sampling pulse (red) and by scanning the THz transient (purple) across it ($t$). The transient is Fourier transformed to obtain the frequency dependent response.

The quantity $\Delta E_R(t,\tau) = E_R^{pumped}(t,\tau) - E_R^{unpumped}(t,\tau)$ was acquired directly at each time delay τ by filtering the electro-optic sampling signal with a lock-in amplifier, triggered by modulation of the mid-infrared pump with a mechanical chopper. This measurement yielded "pump on" *minus* "pump off" reflected electric field.

The differential electric field $\Delta E_R(t,\tau)$ and the stationary reflected electric field $E_R(t)$ were independently Fourier transformed to obtain the complex-valued, frequency dependent $\Delta \tilde{E}_R(\omega)$ and $\tilde{E}_R(\omega)$. In those cases where the pump-induced changes to the reflected field were large enough, we also measured directly $\tilde{E}_R^{pumped}(\omega)$ and calculated the quantity $\Delta \tilde{E}_R(\omega) = \tilde{E}_R^{pumped}(\omega) - \tilde{E}_R^{unpumped}(\omega)$. The two methods yielded identical results. In order to rule out the possibility of phase drifts between $\Delta E_R(t,\tau)$ and $E_R(t)$, we also measured these two quantities at the same time, by simultaneously modulating mid-infrared pump and THz probe and filtering the signal with two lock-in amplifiers[4].

The complex reflection coefficient of the photo-excited sample, $\tilde{r}'(\omega,\tau)$, was determined from the normalized pump-induced changes to the electric field $\Delta\tilde{E}_R(\omega,\tau)/\tilde{E}_R(\omega)$ using the relation

$$\frac{\Delta\tilde{E}_R(\omega,\tau)}{\tilde{E}_R(\omega)} = \frac{\tilde{r}'(\omega,\tau) - \tilde{r}(\omega)}{\tilde{r}(\omega)}$$

where the stationary reflection coefficient $\tilde{r}(\omega)$ was evaluated from the equilibrium optical properties[5] and our own narrow band THz time-domain spectroscopy.

Figure FS2.2A shows the normalized change in the reflected electric field transient measured as a function of sampling delay *t* and at a fixed pump probe time delay (maximum transient response) below and above $T_c$. In Figure FS2.2B we plot the time duration of the light-induced effect at 100 K (above $T_c$), measured by fixing the time delay *t* at the peak of the reflected field transient and scanning the pump-probe time-delay $\tau$. The pump-induced dynamics after excitation is well reproduced by a double exponential decay with time constants $\tau_1 \sim 600$ fs and $\tau_2 \sim 7$ ps (see Fig. FS2.2B). Note that the slow time constant measured here is approximately the same as the one measured below $T_c$ (see Fig. 2 and Fig. 3 in the main text).

From these measurement we extract the changes in the normal-incidence reflectivity as $\frac{\Delta R}{R}(\omega,\tau) = (|\tilde{r}'(\omega,\tau)|^2 - |\tilde{r}(\omega)|^2)/|\tilde{r}(\omega)|^2$ below and above $T_c$ at 10 K and 100 K. They are presented in Figures FS2.2 C.1 and D.1. The measured corresponding phase changes are shown in Figures FS2.2 C.2 and D.2, respectively.

Below $T_c$ strong reflectivity changes appear around 30 cm$^{-1}$. That is the frequency where the JPR is located in the equilibrium reflectivity which is indicated with a black dashed line. We find positive changes at frequencies above the JPR and negative changes at lower frequencies. The corresponding phase change of the transient reflected field evolves sharp around the shifted edge from 0 to approximately $\pi/2$.

Above $T_c$ we observe the appearance of a reflectivity edge at 50 cm$^{-1}$ indicative of the transient high mobility state upon photo-excitation. Also here the corresponding phase change evolves from 0 to approximately $\pi/2$. We note that for a change on a superconduting junction one

---

would expect a $\pi$ shift of the phase. However (see S3), the transient state is an inhomogeneous state of photoexcited regions and an unperturbed background. Therefore the THz transients probe an effective medium for which the phase shift is reduced.

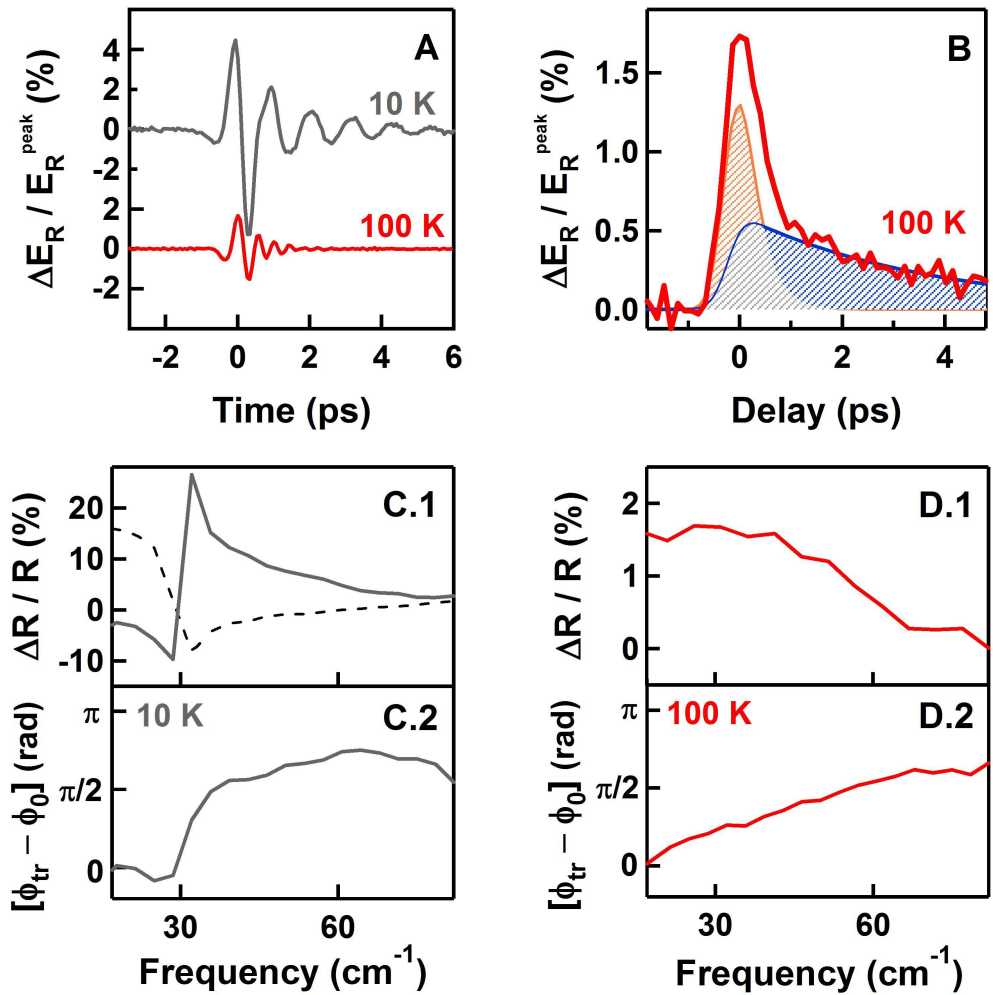

**Figure FS2.2 A.** Differential change in the electric field transient in $YBa_2Cu_3O_{6.5}$ at 10 K (grey) and 100 K (red). **B.** Lifetime dependence of the photo-induced effect. **C.1-D.1.** Frequency dependent differential changes of the reflectivity. **C.2-D.2.** Corresponding phase changes of the transient signal.

These "raw" reflectivity changes require reprocessing. Importantly, the measured changes above $T_c$ are only few percent in size, due to a mismatch between the several-μm penetration depth of the probe and that of the resonant 15-μm pump, which is tuned to the middle of the *reststrahlen* band for this particular phonon and is evanescent over a few hundred nm. At low frequencies, the probe interrogates a volume that is between 10 and 20 times larger than the transformed region beneath the surface, with this mismatch being a function of frequency (see figure FS2.3).

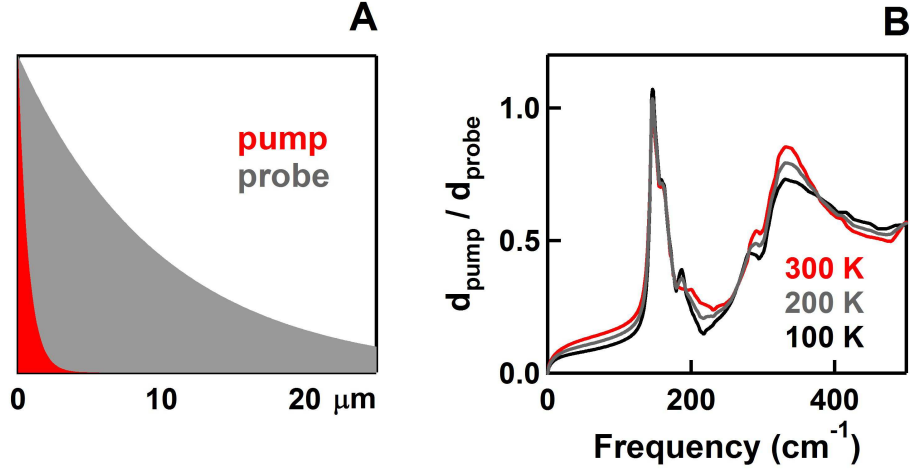

**Figure FS2.3 A.** Schematic of the penetration depth mismatch between resonant vibrational excitation (20-THz frequency or 15-µm wavelength, blue region) and the THz probe (grey region) in $YBa_2Cu_3O_{6.5}$. **B.** Ratio between penetration depths of pump (660 cm$^{-1}$) and probe (20-500 cm$^{-1}$).

To extract the optical constants, the penetration depth mismatch between the mid-infrared pump and the THz probe pulses was taken into account in the data analysis by assuming that the sample was photo-excited homogenously within a thickness corresponding to the mid-infrared penetration depth $d$, and the total reflection coefficient $\tilde{r}$ is a combined contribution from the photo-excited top layer (reflection coefficient $\tilde{r_1}$) and the non-excited bottom layer (reflection coefficient $\tilde{r_0}$)[6]

$$\tilde{r} = \frac{\tilde{r_1} + \tilde{r_0}exp(2i\delta)}{1 + \tilde{r_1}\tilde{r_0}exp(2i\delta)}$$

where $\delta = 2\pi d \, (n_1+ik_1)/\lambda$, and $\lambda$ is the wavelength of the probe pulse. The complex optical constants $\tilde{n}(\omega,\tau)$ of the photo-exited layer were obtained from a numerical solution of the above complex Fresnel equation. For the experiment with the narrowband probe pulses a 30° angle of incidence in the experimental geometry was taken into account.

From the surface refractive index, we calculate the complex conductivity for a volume that is homogeneously transformed,

$$\tilde{\sigma}(\omega,\tau) = \frac{\omega}{4\pi i}[\tilde{n}(\omega,\tau)^2 - \varepsilon_\infty],$$

where $\varepsilon_\infty = 4.5$, a standard value for cuprates[7].

---

[6] M. Dressel and G. Grüner, *Electrodynamics of Solids,* Cambridge University Press, Cambridge (2002).

[7] D. van der Marel, H. J. A. Molegraaf, J. Zaanen, et al., "Quantum critical behaviour in a high-$T_c$ superconductor," Nature **425**, 271 (2003).

In figure 5 in the main text the changes in the reflectivity, $\frac{\Delta R}{R}(\omega, \tau) = (R(\omega, \tau) - R_{eq}(\omega))/R_{eq}(\omega)$, are recalculated by assuming normal-incidence reflection,

$$R(\omega) = \left|\frac{1 - \tilde{n}(\omega)}{1 + \tilde{n}(\omega)}\right|^2.$$

**S3. Fits to the optical properties**

All transient optical properties can be well explained by Bruggeman's effective medium model[8],

$$f \frac{\tilde{\varepsilon}_T(\omega) - \tilde{\varepsilon}_E(\omega)}{\tilde{\varepsilon}_T(\omega) + 2\tilde{\varepsilon}_E(\omega)} + (1 - f) \frac{\tilde{\varepsilon}_{eq}(\omega) - \tilde{\varepsilon}_E(\omega)}{\tilde{\varepsilon}_{eq}(\omega) + 2\tilde{\varepsilon}_E(\omega)} = 0,$$

where the effective medium dielectric function $\tilde{\varepsilon}_E(\omega)$ is determined by the dielectric functions $\tilde{\varepsilon}_T(\omega)$ of the photo-susceptible region, and by the dielectric function $\tilde{\varepsilon}_{eq}(\omega)$ of the unperturbed ground state. Here Munzar's multi-layer model[9] is used to calculate $\tilde{\varepsilon}_T(\omega)$ and $\tilde{\varepsilon}_{eq}(\omega)$. Fig.FS3.1 a1-a4 shows the effective medium fit for the maximum transient response, in which a transformed volume fraction of $f$ = 20% is used. All the transient changes are well reproduced by the fit, for example, the mixing of two superconductors (with different inter-bilayer Josephson plasma frequencies $\omega_{Jp1}$ and $\omega'_{Jp1}$) in real space perfectly explain the plateau observed around 30 cm$^{-1}$ in $\sigma_2$ (Fig.S3.1a1), and the small peak in $\sigma_1(\omega)$ around 50 cm$^{-1}$ (Fig.S3.1a3).

The optical constants of the two superfluids are shown in Fig.FS3.1 b1-b4. The blue curves in Fig.FS3.1b show the part of the sample which keeps the equilibrium properties ($\omega_{Jp1}$=110 cm$^{-1}$, $\omega_{Jp2}$ = 1030 cm$^{-1}$, 320 cm$^{-1}$ phonon width = 40 cm$^{-1}$) [10], and the red curves show the optical properties of the photo-susceptible region ($\omega'_{Jp1}$=310 cm$^{-1}$ and $\omega'_{Jp2}$ = 950 cm$^{-1}$, 320 cm$^{-1}$ phonon width = 4 cm$^{-1}$).

---

[8] Tuck C. Choy, *Effective medium theory: principles and applications.* Oxford University Press, New York, (1999).

[9] D. Munzar, C. Bernhard, A. Golnik, et al, "Anomalies of the infared-active phonons in underdoped YBCO as an evidence for the intra-bilayer Josephson effect," *Solid State Commun.* **112,** 365 (1999).

[10] The position of the Josephson plasma edges in the reflectivity, and the peaks in the loss function, are located at frequencies much smaller than $\omega_{Jp1}$ and $\omega_{Jp2}$. Here the edge/peak positions are determined by the screened plasma frequency $\widetilde{\omega_{Jp}} = \omega_{Jp}/\sqrt{\varepsilon_\infty}$. The interband contribution shifts the peak/edge positions further to even lower frequencies than $\widetilde{\omega_{Jp}}$.

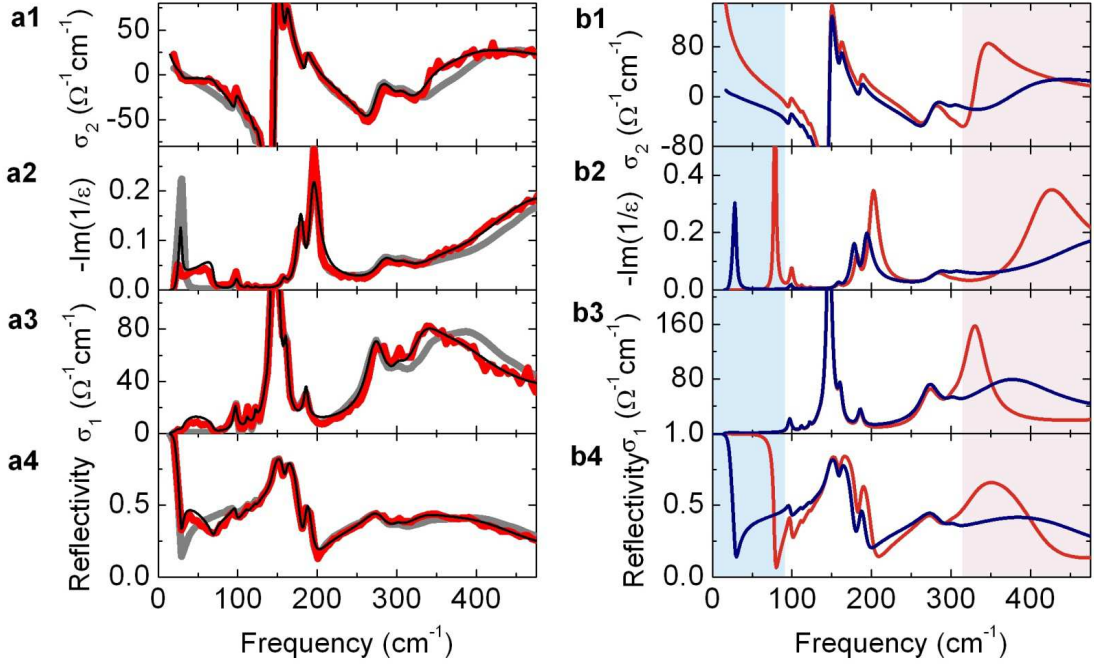

**Figure FS3.1** Effective medium fit for the transient optical properties below Tc. **(a1-a4)**: equilibrium (gray), photo-excited (red), and effective medium fit (black) of the optical properties of $YBa_2Cu_3O_{6.5}$ at the maximum transient response. **(b1-b4)**: Two components used for the effective medium fit. For 20% of the photo-excited volume fraction (red), the inter-bilayer Josephson plasma frequency increases, which results in an enhanced low frequency $\sigma_2$, and a blue-shift of the inter-bilayer Josephson plasma mode from 30 to 80 cm$^{-1}$ in the loss function and reflectivity. In the high frequency region, a reduction of the intra-bilayer Josephson plasma frequency results in a red-shift of the transverse plasma mode in $\sigma_1$, a corresponding spectral weight transfer back to the planar oxygen phonon, and a damping reduction for this phonon. 80% of the photo-excited volume retains the equilibrium properties (blue).

The transient data in the normal state (at 60 K) displays strong resemblance to the below $T_c$ data, and can be analyzed within a similar effective medium model with a filling factor of 19%. (Fig.FS3.2). The equilibrium properties are fitted with $\omega_{Jp1}$=0 cm$^{-1}$, $\omega_{Jp2}$ = 1030 cm$^{-1}$, and the photo-susceptible properties are fitted with $\omega'_{Jp1}$=250 cm$^{-1}$ and $\omega'_{Jp2}$ = 960 cm$^{-1}$ using Munzar's multi-layer model[9].

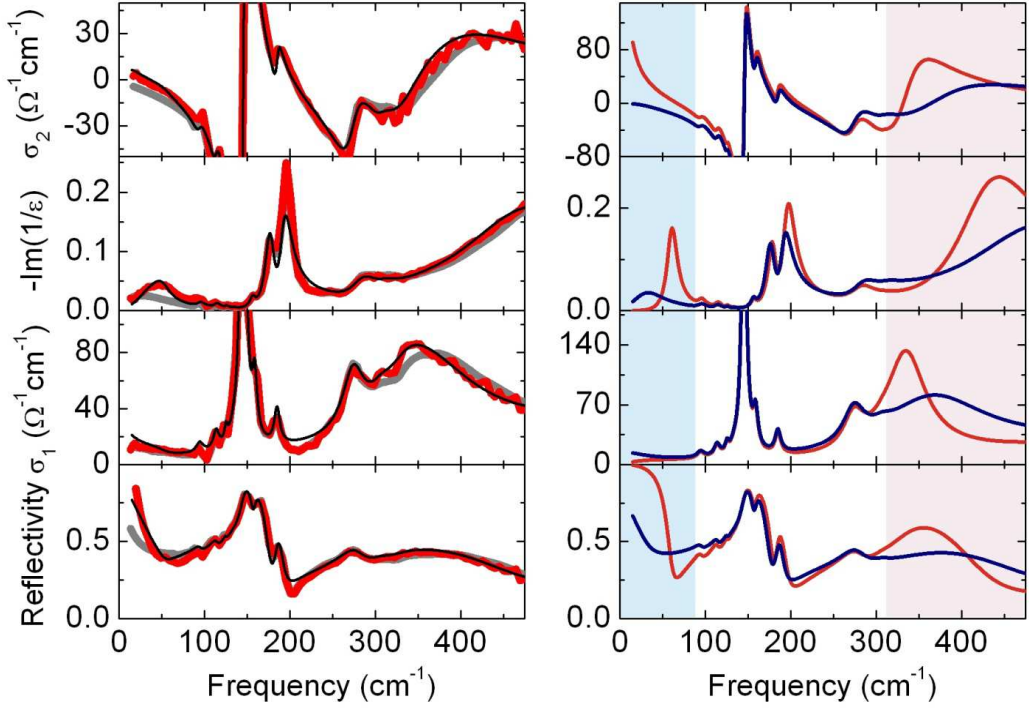

**Figure FS3.2** Effective medium fit for the transient optical properties. Left panels: equilibrium (gray), photo-excited (red), and effective medium fit (black) of the optical properties of $YBa_2Cu_3O_{6.5}$ at the maximum transient response. Right panels: the two components used for the effective medium fit. A new inter-bilayer Josephson plasma mode is developed and a red-shift of the transverse plasma mode in seen in 19% of the photo-excited volume.

To calculate the two components $\tilde{\varepsilon}_T(\omega)$ and $\tilde{\varepsilon}_{eq}(\omega)$ of the effective medium fit, we also use the conventional Drude-Lorentz model[6]. As shown in Fig. FS3.3, both Munzar's multi-layer model and the Drude-Lorentz model give quantitatively comparable results for the effective medium fit.

For the conventional Drude-Lorentz fit of the below $T_c$ data, we use $\omega_{Jp1}$ = 127 cm$^{-1}$ to fit the low frequency Josephson plasma edge, and one Lorentz term centered at 376 cm$^{-1}$ to fit the 400 cm$^{-1}$ mode in $\sigma_1$. For the dielectric function $\tilde{\varepsilon}_T(\omega)$, $\omega'_{Jp1}$=392 cm$^{-1}$ is used to fit the enhanced superconducting response at low frequencies, and the red-shifted 400 cm$^{-1}$ mode and the corresponding spectral weight redistribution is fitted by a Lorentz oscillator at 330 cm$^{-1}$ with a reduced width from 144 to 35 cm$^{-1}$.

For the above $T_c$ data, we use a Drude term with $\sigma_{dc}$=14 Ω$^{-1}$cm$^{-1}$ and a scattering rate of Γ=25.5 cm$^{-1}$ to fit the small metallic background at low frequency. A Lorentz term at 368 cm$^{-1}$ is used to fit the equilibrium transverse plasma mode in $\sigma_1$. For $\tilde{\varepsilon}_T(\omega)$, we replace the low frequency Drude contribution with an inter-bilayer Josephson plasma mode $\omega'_{Jp1}$=297

cm$^{-1}$, and shift the Lorentz peak from 368 to 335 cm$^{-1}$ and reduce its width from 151 to 55 cm$^{-1}$ to fit the red-shift of the transverse plasma mode.

The advantage of Munzar's multi-layer model is that no additional term is needed to fit the 400 cm$^{-1}$ mode, which comes out directly from the bilayer structure. Its frequency is mainly determined by the intra-bilayer plasma frequency $\omega_{Jp2}$, and its spectral weight is strongly coupled to the 320 cm$^{-1}$ planar-oxygen phonon. The light-induced red-shift and the reshaping of the 400 cm$^{-1}$ mode can be directly explained by a reduction of the intra-bilayer plasma frequency $\omega'_{Jp2}$. The sharpening of the red-shifted peak in $\sigma_1$ is results from a reduction of the 320-cm$^{-1}$ phonon line width.

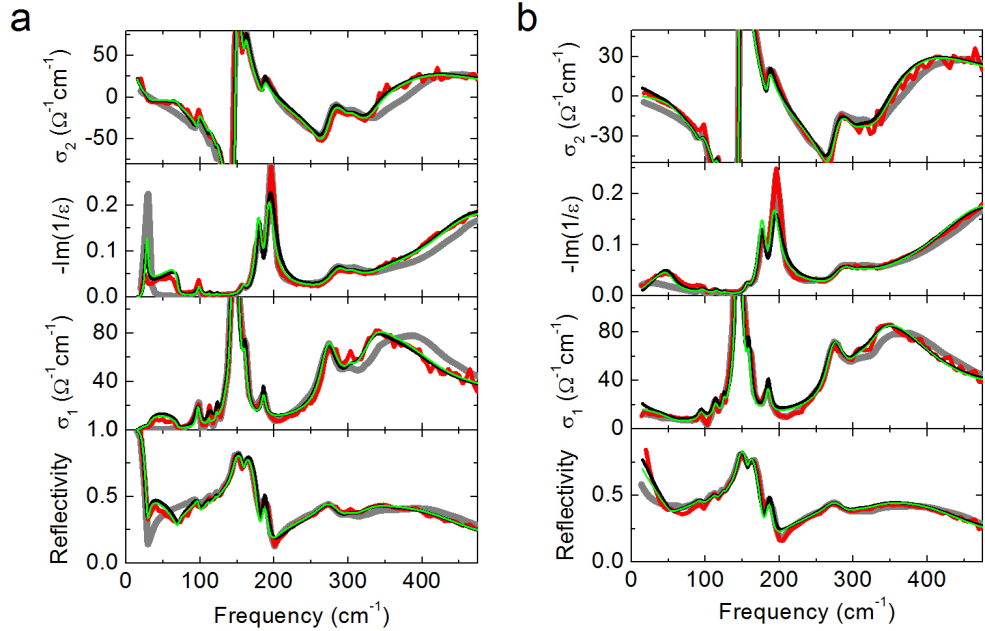

**Figure FS3.3.** Effective medium fit for the transient optical properties (a) below $T_c$ ($T$ = 10 K), (b) above $T_c$ ($T$ = 60 K). Equilibrium (gray), photo-excited (red), and effective medium fits from the Drude-Lorentz model (black) and from Munzar's multilayer model (green) for YBa$_2$Cu$_3$O$_{6.5}$.

Full fits to YBa$_2$Cu$_3$O$_{6.5}$ narrowband data above $T_c$ (Fig. 5 in the main text) are shown in Figure FS3.4. Letting only the filling fraction of the high mobility state as free parameter allows describing the data at all temperatures. The photo-induced response can be fit equally well by assuming a perfect conductor with 7 ps lifetime or with infinitely long lifetime. If one assumes instead a carrier scattering time much shorter than 7 ps, for instance $\tau_s$=1 ps, no agreement with the experimental results can be found. As shown in Figure FS3.5a, a conductor with $\tau_s$ = 1 ps and $\omega_{Jp1}$~50 cm$^{-1}$ would display, unlike the measured transient state, a significant increase in $\sigma_1(\omega)$ over the whole measured range, no low-frequency divergence in $\sigma_2(\omega)$, and no clear reflectivity edge.

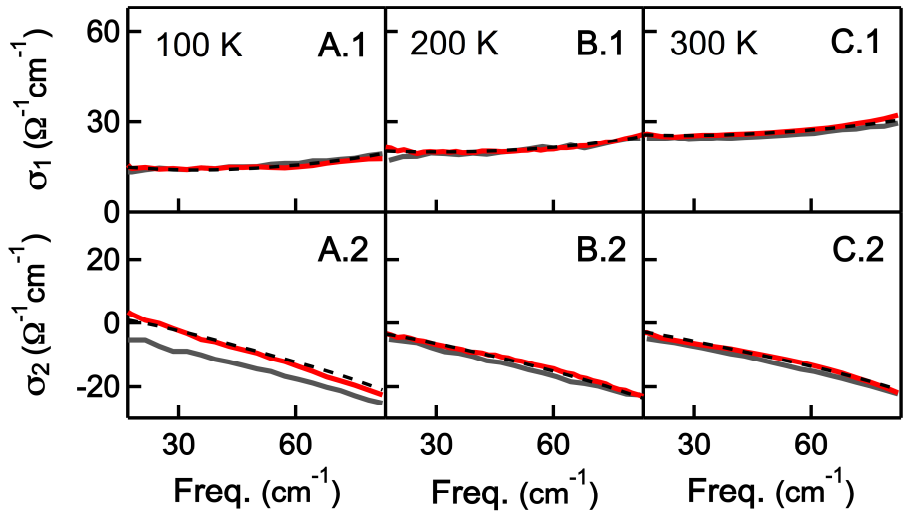

**Figure FS3.4**: Transient optical properties of $YBa_2Cu_3O_{6.5}$ at maximum transient response (red lines) at three temperatures: **(A)** 100 K, **(B)** 200 K, and **(C)** 330 K. The volume fraction of coherent conductor was extracted from effective medium fits to the optical properties (black dashed lines).

The lifetime of the transient state decreases with increasing temperature, down to ~2 ps at room temperature. We find that the transient optical properties remain consistent with a state in which its scattering rate is limited only by the lifetime of the state itself.

Also a small increase (< 10%) in the $\sigma_1(\omega)$ of the normal state conductivity was necessary for the fit. Such a small Ohmic component suggests that incoherent carriers are also being excited.

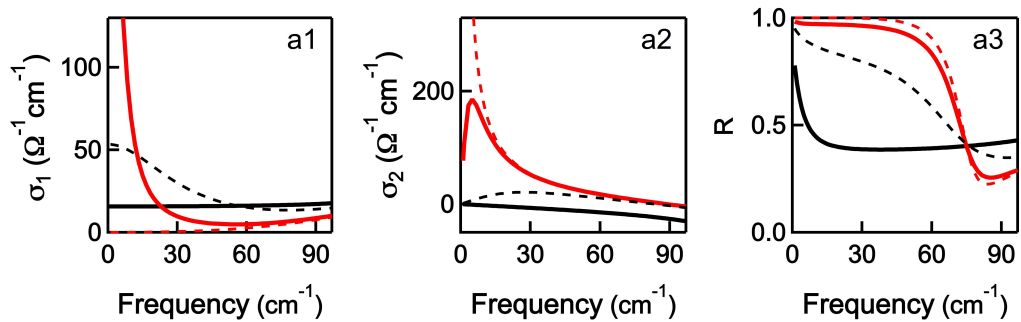

**Figure FS3.5 (a1-a3)** Optical properties of the equilibrium material (black lines), a conductor with infinite (red dashed line), 7 ps (red lines), and 1 ps (black dashed line) carrier scattering time, calculated with the Drude-Lorentz model.

**S4 Wavelength dependence**

Figure FS4.1 shows the pump wavelength dependence of the 1-THz imaginary conductivity in photo-excited $YBa_2Cu_3O_{6.5}$. This quantity, which is proportional to the strength of the transient high mobility state, follows the apical oxygen phonon resonance (dashed curve), and correlates

almost exactly with it when the line shape is convolved with the pump pulse bandwidths (blue line). Note that measurements could not be performed at longer wavelengths than 16 μm with our optical parametric amplifier.

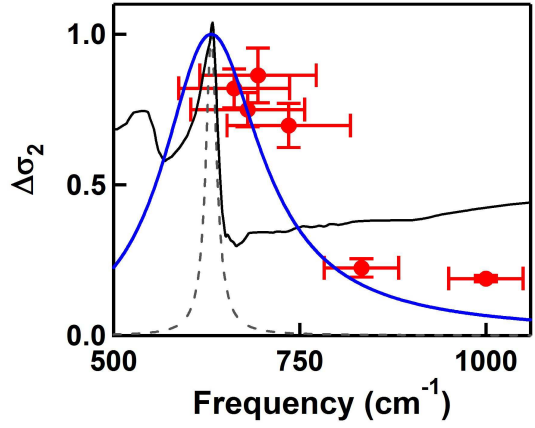

**Figure FS4.1**: Normalized $\Delta\sigma_2$ at 1 THz plotted as a function of pump wavelength (red dots) and measured at constant pump fluence. The horizontal bars indicate the pump spectral linewidth. The black line describes the equilibrium optical conductivity of $YBa_2Cu_3O_{6.5}$. The resonant frequency of the apical oxygen mode is indicated in the dashed black curve, and in the blue curve when convolved with the pump pulse frequency bandwidth.

**S5 Evaluation of the peak lattice distortion**

For these electric fields, the amplitude of the field-induced atomic displacement due to mid-infrared photo-excitation can be estimated assuming an ionic bonding between the apical $O^{2-}$ ion and the Cu(1) ion in the CuO chains.

The atomic polarizability can be derived as $\boldsymbol{P}(\omega_0) = \varepsilon_0 \chi(\omega_0) \boldsymbol{E}(\omega_0)$, where $\omega_0 = 20$ THz and $\|\boldsymbol{E}(\omega_0)\| \cong 3$ MV/cm. The susceptibility $\chi(\omega_0)$ is calculated from the *c*-axis equilibrium optical conductivity using $|\varepsilon_0 \chi(\omega_0)| = \left|\frac{\sigma(\omega_0)}{\omega_0}\right|$. As $|\sigma_1(\omega_0)| \cong |\sigma_2(\omega_0)| \cong 30\ \Omega^{-1}\mathrm{cm}^{-1}$ (from data in Refs. 5 and 6), one gets $|\varepsilon_0 \chi(\omega_0)| \cong 2 \cdot 10^{-12} \Omega^{-1} \mathrm{cm}^{-1} \mathrm{s}$ and $\|\boldsymbol{P}(\omega_0)\| \cong 6 \cdot 10^{-6}\ \mathrm{C} \cdot \mathrm{cm}^{-2}$.

The average size of the photo-induced electric dipole, i.e., the displacement of the oxygen ions, is then given by $d = \|\boldsymbol{P}(\omega_0)\|/nQ$, where $n$ is the density of dipoles (2 per unit cell of volume 173 Å$^3$) and $Q = 3e$. This yields $d \sim 10$ pm, which is approximately 5% of the equilibrium Cu-O distance.

**REFERENCES (Main text)**